\documentclass{ws-mpla}
\usepackage[super]{cite}
\usepackage{graphicx}
\usepackage[utf8x]{inputenc}
\usepackage[english]{babel}
\usepackage{graphicx}
\usepackage{float}
\usepackage{multirow}
\usepackage{cite}
\usepackage{setspace, longtable, amstext}
\usepackage{selinput}
\usepackage{hyperref}

\usepackage{amssymb}
\usepackage{amsmath}

\usepackage{amsopn}
\def\beq{\begin{equation}}
\def\eeq{\end{equation}}
\def\bea{\begin{eqnarray}}
\def\eea{\end{eqnarray}}

\begin{document}

\markboth{A. Amekhyan, S. Sargsyan, A. Stepanian}
{Galactic sparsity and $\Lambda$-gravity}

\catchline{}{}{}{}{}

\title{Galactic sparsity and $\Lambda$-gravity}

\author{A. Amekhyan}
\author{S. Sargsyan}
\author{A. Stepanian}
\address{Center for Cosmology and Astrophysics, Alikhanian National Laboratory, Yerevan, Armenia
}

\maketitle

\begin{history}
\received{... 2020}
\accepted{...}
\end{history}

\begin{abstract}
The sparsity parameter for clusters of galaxies is obtained in the context of $\Lambda$-gravity. It is shown that, the theoretical estimated values are within the reported error limits of the measured data. Thus, in the future the sparsity parameter can serve as an informative new test to detect the discrepancy between General Relativity and $\Lambda$-gravity.  
\end{abstract}


\ccode{PACS numbers: 98.80.}

\section{Introduction}	

Newton's theorem on the equivalence of gravity of spheres and point masses provides a unifying picture for describing the dark sector (DS)\cite{G,GS1,GS2}. Namely, within this approach, on the one hand, the cosmological constant $\Lambda$ is entered in the equations of General Relativity (GR) and explains the accelerated expansion of the Universe as the best candidate for dark energy (DE), on the other hand, it enables one to describe the dynamics of dark matter (DM) in astrophysical configurations \cite{GS1}. As a result, the weak-field modification of GR is given by the following metric
\begin {equation} \label{mod}
g_{00} = 1 - \frac{2 G m}{c^2r} - \frac{\Lambda r^2}{3}\,; 
\qquad g_{rr} = \left(1 - \frac{2 G m}{c^2r} - \frac {\Lambda r^2}{3}\right)^{-1}\,.
\end {equation} 

It is worth mentioning that although the above metric was known before as the Schwarzschild-de Sitter solution, which was used to describe the Universe when a single spherically symmetric object is immersed in de Sitter background, its deduction based on Newton's theorem can be used to provide a description for astrophysical structures such as galaxy clusters within the weak-field limit of GR \cite{GS2}. Namely, within this approach, the weak-field of all objects contains the additional $\Lambda$ term according to Eq.(\ref{mod}).Thus, $\Lambda$-gravity assumes that $\Lambda$ with its current numerical value \cite{Pl}, is a fundamental constant of nature \cite{GS4} which naturally enters in both relativistic and weak-field gravity equations.

Indeed, the general function for force $\mathbf{F}(r)$ satisfying Newton's theorem used in the above metric has the following form (see \cite{G,GS1,G1}) 
\begin{equation}\label{FandU}
\mathbf{F}(r) = \left(-\frac{A}{r^2} + Br\right)\hat{\mathbf{r}}\,
\end{equation}     
in which the first term is associated to the ordinary Newtonian gravity, while the linear term is related to the contribution of $\Lambda$.

The appearance of the linear term in Eq.(\ref{FandU}) and its consideration as $\Lambda$ in Eq.(\ref{mod})  has a clear group-theoretical background \cite{GS1}.  Namely, depending on the sign of $\Lambda$ - being positive, negative or zero - the corresponding vacuum solutions for GR equations and their isometry groups are obtained according to Table 1.

\begin{table}\label{tab1} 
\centering
\tbl{The vacuum solutions of GR}
{
\begin{tabular}{ |p{0.8cm}||p{2.7cm}|p{2.1cm}|p{1.5cm}|}
\hline
\multicolumn{4}{|c|}{Background geometries} \\
\hline
Sign& Spacetime&Isometry group&Curvature\\
\hline
$\Lambda > 0$ &de Sitter (dS) &O(1,4)&+\\
$ \Lambda = 0$ & Minkowski (M) & IO(1,3)&0 \\
$\Lambda <0 $ &Anti de Sitter (AdS) &O(2,3)&-\\
\hline
\end{tabular}
}
\end{table}
\noindent 
Then, the Lorentz group O(1,3) is the stabilizer group for all of these maximally symmetric Lorentzian 4D-geometries. The group O(1,3) of orthogonal transformations implies a spherical symmetry (in Lorentzian sense) at each point of these geometries  
\begin{equation}\label{qspace}
dS= \frac{O(1,4)}{O(1,3)}\,, \quad M=\frac{IO(1,3)}{O(1,3)}\,, \quad AdS=\frac{O(2,3)}{O(1,3)} \,.
\end{equation}
In the non-relativistic limit the full Poincare group IO(1,3) is reduced to Galilei group Gal(4)=(O(3)$\times$R)$\ltimes$R${}^{6}$, as an action of O(3)$\times$R  on group of boosts and spatial translations R${}^{6}$. Similarly, for non-relativistic limit of O(1,4) and O(2,3) groups one has
\begin{equation}\label{nrel}
O(1,4) \to (O(3) \times O(1,1)) \ltimes R^6\,,\quad O(2,3) \to (O(3) \times O(2) ) \ltimes R^6\,.
\end{equation}
Consequently, the Galilei spacetime is achieved via quotienting Gal(4) by O(3)$\times$R${}^{3}$, while the Newton-Hooke NH${}^{\pm}$ spacetimes are given by the same quotient group but for groups achieved in Eq.(\ref{nrel}). For all of these cases, O(3) is the stabilizer group of spatial geometry, that is each point (in spatial geometry) admits O(3) symmetry. This statement can be regarded as the group-theoretical formulation of Newton's theorem.
 
The next important fact is that, the force in the form of Eq.(\ref{FandU}) defines a non-force-free field inside a spherical shell, thus drastically contrasting with Newton's gravity when the shell has force-free field in its interior. Thus by considering the galactic halos in the context of Newtonian gravity no force can influence the interior region. But this is somehow problematic with the nature of Newtonian gravity as the observations indicate that galactic halos do determine features of galactic disks \cite{Kr}. However, comparing to the Newtonian gravity, the $\Lambda$ term in Eq.(\ref{mod}) creates a non-zero force inside the shell. Moreover, the $\Lambda$-gravity is able to describe the observational features of galactic halos \cite{G,Ge}, of groups and clusters of galaxies \cite{GS2}.

In what follows, we continue the same path of analysis, by studying the so-called sparsity parameter for clusters of galaxies in the context of $\Lambda$-gravity.

\section{DM and virial theorem}

As widely known, the dark matter indications for galactic systems arise when using the virial theorem, $2K+W=0$, where $K=\frac{1}{2}\sum_{i=1}^{N} m_{i}v_{i}^{2}$ and $W=\frac{-G}{2}\sum_{i=1}^{N}\sum_{j\neq i}\frac{m_{i}m_{j}}{\|\vec{r_{i}}-\vec{r_{j}}\|}$ are total kinetic and potential energies of a self gravitating, stationary system respectively. According to currently favored large-scale structure formation mechanism, dark matter must be made up of cold, non-relativistic (massive) particles. Moreover, the candidates of DM particles are electrically neutral, long-lived and interacting only via gravitation. The popular candidates of such particles are Weakly Interacting Massive Particles (WIMPs) \cite{Schumann} or axions \cite{Knirck}. However there are a few ``ordinary baryonic" candidates, which are Massive Astrophysical Compact Halo Objects (MACHOs) \cite{Yang} or interstellar/intergalactic medium. It is also possible to explain the DM puzzle via modified gravity \cite{Sharma, GS4} theories.
Even without having any knowledge about DM composition, it is still possible to study its spatial distribution via numerical simulations, such as the Navarro-Frenk-White (NFW) \cite{NFW} distribution profile
\begin{equation}
\frac{\rho_{r}}{\rho_{crit}}=\frac{\delta_{0}}{(r/r_{s})(1+r/r_{s})^2},
\end{equation}
where $\rho_{crit}=\frac{3H^{2}}{8 \pi G}$ is the critical density and $r_{s}$ is characteristic radius. The $\delta_{0}$ is called the overdensity parameter and defined as
\begin{equation}
\delta_{0}=\frac{200}{3}\frac{n^{3}}{(ln(1+n)-n/(1+n))},
\end{equation} 
in which $n$ is the concentration of particles. The dark matter mass inside the radius $r$ is
\begin{equation}
M(r)=4\pi \rho_{crit}\delta_{0}r_{s}^{3}\left[\frac{1}{1+nr_{s}}-1+ln(1+nr_{s})\right].
\end{equation}
Here $r_{s}=r/R_{200}$ and $R_{200}$ is the virial radius defined as the radius at which the density of overdense region (where virial equilibrium holds) is $\rho=200\rho_{crit}$.
Another possible source to study DM is Cosmic Microwave Background (CMB) temperature asymmetry, founded in galactic halos \cite{VG, FDP}. The reason of this asymmetry is Doppler effect
\begin{equation}
\frac{|\Delta T|}{T_{CMB}}=\frac{2vsini}{c}\tau,
\end{equation} 
where $\Delta T$ is CMB temperature asymmetry, $v$ is galactic rotational velocity, $i$ is galaxy inclination angle and $\tau$ is optical depth of the interstellar medium (ISM). In this way one can examine the possible contribution of ISM in the galactic dark halos \cite{AA}.

\section{Sparsity}
Before starting our analysis it should be noticed that in the context of $\Lambda$-gravity the virial theorem is written as
\begin{equation}\label{VirL}
\sigma^2 = \frac{GM}{r} - \frac{\Lambda c^2 r^2}{3}
\end{equation}
where $\sigma$ is regarded as the velocity dispersion.

Recently, the sparsity parameter has been used to study different modified theories of gravity \cite{Spar}. It is defined, based on virial theorem, as follows

\begin{equation}\label{Sp}
s_{(c_1, c_2)} = \frac{M_{c_1}}{M_{c_2}}
\end{equation}
Here, $M_{c_i}$ is the total mass enclosed in the radius $r_{c_i}$ within which the total density is $c_i$ times denser than the critical density
\begin{equation}\label{Mass}
M_{c_i} = \frac{4 \pi r_{c_i}^3}{3} c_i \rho_{crit}
\end{equation}
Considering the virial theorem and Eq.(\ref{mod}) simultaneously, we can obtain the sparsity parameter in the context of $\Lambda$-gravity, which is
\begin{equation}\label{SpL}
s_{\Lambda(c_1, c_2)} = \frac{M_{c_1} - \frac{2 \Lambda c^2 M_{c_1}}{3 c_1 H^2}}{M_{c_2} - \frac{2 \Lambda c^2 M_{c_2}}{3 c_2 H^2}}
\end{equation}
Clearly, by comparing to the sparsity parameter in the context of standard GR/Newtonian gravity, it turns out that if we put $\Lambda=0$, the result of Eq.(\ref{Sp}) will be recovered. Namely, the Eq.(\ref{SpL}) shows that in the context of $\Lambda$-gravity, due to the presence of $\Lambda$ term, some new additional terms are appeared. Thus, in order to preserve the consistency of $\Lambda$-gravity with observational results, it is essential to check whether the numerical value of $s_{\Lambda(c_1, c_2)}$ lies inside the error limits of $s$ or not.

Here we use the data of CLASH survey \cite{CLASH}, to compare the sparsity parameters of the reported data with the ones of $\Lambda$-gravity. The results have been illustrated in Tables 2-4.

Comparing the sparsity parameter with $s_{\Lambda}$, it turns out that for all cases, $s_{\Lambda}$ lies inside the error limits of $s$. Such result, in its turn, shows that there is no inconsistency between the observational results and the predictions of $\Lambda$-gravity. Moreover, it ensures that we can use the sparsity parameter as a new criterion to detect the deviations from GR in future.  

\begin{table}\label{table2}
\tbl{Sparsity parameter $s_{(200, 500)}$ for the reported data and $\Lambda$-gravity}
{
\centering
\begin{tabular}{ |p{2.4cm}|p{1.5cm}|p{2.1cm}|p{2.1cm}|p{1.5cm}|p{1.5cm}| }
\hline
Cluster & redshift $z$ & $M_{500}$ $(10^{15}M_{\odot})$ &$M_{200}$ $(10^{15}M_{\odot})$ & $s_{(200, 500)}$& $s_{{\Lambda (200, 500)}}$\\
\hline
Abell 383& 0.188 & $0.61\pm	0.07$&	$0.87\pm	0.07$& $1.426^{0.314}_{0.249}$ & 1.420\\
Abell 209& 0.206 & $0.63\pm	0.07$&	$0.95\pm	0.07$& $1.507^{0.313}_{0.250}$&1.501\\
				
Abell 2261&	0.225 & $0.95\pm	0.16$&	$1.42\pm	0.17$& $1.494^{0.517}_{0.368}$&1.488\\
RXJ2129+0005& 0.234 & $0.43\pm	0.04$&	$0.61\pm	0.06$& $1.418^{0.299}_{0.248}$&1.412\\
Abell 611&	0.288 & $0.57\pm	0.04$&	$0.85\pm	0.05$& $1.491^{0.206}_{0.179}$&1.485\\
MS2137-2353& 0.313 & $0.68\pm	0.05$&	$1.04\pm	0.06$& $1.529^{0.216}_{0.186}$&1.523\\
RXCJ2248-4431& 0.348 & $0.76\pm	0.12$&	$1.16\pm	0.12$& $1.526^{0.473}_{0.344}$&1.519\\
MACSJ1115+0129& 0.352 & $0.54\pm	0.08$&	$0.90\pm	0.09$& $1.666^{0.485}_{0.360}$&1.659\\
MACSJ1931-26& 0.352 & $0.45\pm	0.04$&	$0.69\pm	0.05$& $1.533^{0.271}_{0.227}$&1.526\\
RXJ1532.8+3021&	0.363 & $0.34\pm	0.08$&	$0.53\pm	0.08$& $1.558^{0.787}_{0.487}$&1.552\\
MACSJ1720+3536&	0.391 & $0.53\pm	0.08$&	$0.75\pm	0.08$& $1.415^{0.429}_{0.316}$&1.409\\
MACSJ0429-02& 0.399 & $0.53\pm	0.13$&	$0.80\pm	0.14$& $1.509^{0.840}_{0.509}$&1.503\\
MACSJ1206-08&	0.439 & $0.60\pm	0.11$&  $0.86\pm	0.11$& $1.433^{0.546}_{0.376}$&1.427\\
MACSJ0329-02&	0.450 & $0.50\pm	0.09$&	$0.73\pm	0.10$& $1.460^{0.564}_{0.392}$&1.453\\
RXJ1347-1145&	0.451 & $0.79\pm	0.19$&	$1.16\pm	0.19$& $1.468^{0.78}_{0.47}$&1.462\\
MACSJ1311-03&	0.494 & $0.32\pm	0.19$&  $0.46\pm	0.03$& $1.437^{2.331}_{0.594}$&1.431\\
MACSJ1423+24&	0.545 & $0.41\pm	0.06$&  $0.57\pm	0.10$& $1.390^{0.524}_{0.390}$&1.384\\
MACSJ0744+39&	0.686 & $0.49\pm	0.04$&	$0.70\pm	0.04$& $1.428^{0.215}_{0.183}$&1.422\\
CLJ1226+3332&	0.890 & $1.08\pm	0.09$&	$1.56\pm	0.10$& $1.444^{0.232}_{0.196}$&1.438\\
\hline
\end{tabular}
}
\end{table}

\begin{table}\label{tab3}
\tbl{The same as Table 2, but for $s_{(200, 2500)}$}
{
\centering
\begin{tabular}{ |p{2.4cm}|p{2.4cm}|p{2.1cm}|p{1.5cm}|p{1.5cm}| }
\hline

Cluster & $M_{2500}$ $(10^{15}M_{\odot})$ & $M_{200}$ $(10^{15}M_{\odot})$& $s_{(200, 2500)}$ & $s_{\Lambda (200, 2500)}$\\ \hline
\hline
Abell 383&	$0.26\pm0.05$&	$0.87\pm	0.07$& $3.34^{1.13}_{0.76}$&3.33\\
Abell 209&	$0.22\pm0.05$&	$0.95\pm	0.07$& $4.31^{1.68}_{1.05}$&4.30\\
				
Abell 2261&	$0.34\pm0.12$&	$1.42\pm	0.17$& $4.17^{3.05}_{1.45}$&4.15\\
RXJ2129+0005&$0.18\pm0.03$&	$0.61\pm	0.06$& $3.38^{1.07}_{0.76}$&3.37\\
Abell 611&	$0.21\pm0.04$&	$0.85\pm	0.05$& $4.04^{1.24}_{0.84}$&4.03\\
MS2137-2353&$0.23\pm0.04$&	$1.04\pm	0.06$& $4.52^{1.26}_{0.89}$&4.50\\
RXCJ2248-4431&$0.27\pm0.07$&$1.16\pm	0.12$& $4.29^{2.10}_{1.23}$&4.27\\
MACSJ1115+0129&$0.15\pm0.05$& $0.90\pm	0.09$& $6.00^{3.90}_{1.95}$&5.97\\
MACSJ1931-26&	$0.16\pm0.03$&	$0.69\pm	0.05$& $4.31^{1.37}_{0.94}$&4.29\\
RXJ1532.8+3021&	$0.11\pm0.05$&	$0.53\pm	0.08$& $4.81^{5.34}_{2.00}$&4.79\\
MACSJ1720+3536&	$0.22\pm0.06$&	$0.75\pm	0.08$& $3.40^{1.77}_{1.01}$&3.39\\
MACSJ0429-02&	$0.19\pm0.11$&	$0.80\pm	0.14$& $4.21^{7.53}_{2.01}$&4.19\\
MACSJ1206-08&	$0.25\pm0.08$&  $0.86\pm	0.11$& $3.44^{2.26}_{1.16}$&3.42\\
MACSJ0329-02&	$0.20\pm0.06$&	$0.73\pm	0.10$& $3.65^{2.27}_{1.22}$&3.63\\
RXJ1347-1145&	$0.31\pm0.13$&	$1.16\pm	0.19$& $3.74^{3.75}_{1.53}$&3.72\\
MACSJ1311-03&	$0.14\pm0.02$&  $0.46\pm	0.03$& $3.28^{0.79}_{0.59}$&3.27\\
MACSJ1423+24&	$0.18\pm0.08$&  $0.57\pm	0.10$& $3.16^{3.53}_{1.35}$&3.15\\
MACSJ0744+39&	$0.20\pm0.03$&	$0.70\pm	0.04$& $3.49^{0.85}_{0.63}$&3.48\\
CLJ1226+3332&	$0.43\pm0.07$&	$1.56\pm	0.10$& $3.62^{0.98}_{0.70}$&3.61\\
\hline
\end{tabular}
}
\end{table}

\begin{table}\label{tab4}
\tbl{The same as Table 2, but for $s_{(500, 2500)}$}
{
\centering
\begin{tabular}{ |p{2.4cm}|p{2.4cm}|p{2.4cm}|p{1.5cm}|p{1.5cm}| }
\hline

Cluster & $M_{2500}$ $(10^{15}M_{\odot})$& $M_{500}$ $(10^{15}M_{\odot})$& $s_{(500, 2500)}$ & $s_{\Lambda (500, 2500)}$\\ \hline
\hline
Abell 383&	$0.26\pm0.05$&	$0.61\pm	0.07$& $2.34^{0.89}_{0.60}$&2.33\\
Abell 209&	$0.22\pm0.05$&	$0.63\pm	0.07$& $2.86^{1.25}_{0.78}$&2.85\\
				
Abell 2261&	$0.34\pm0.12$&	$0.95\pm	0.16$& $2.79^{2.25}_{1.07}$&2.78\\
RXJ2129+0005&$0.18\pm0.03$&	$0.43\pm	0.04$& $2.38^{0.74}_{0.53}$&2.37\\
Abell 611&	$0.21\pm0.04$&	$0.57\pm	0.04$& $2.71^{0.87}_{0.59}$&2.70\\
MS2137-2353&$0.23\pm0.04$&	$0.68\pm	0.05$& $2.95^{0.88}_{0.62}$&2.94\\
RXCJ2248-4431&$0.27\pm0.07$&$0.76\pm	0.12$& $2.81^{1.58}_{0.93}$&2.80\\
MACSJ1115+0129&$0.15\pm0.05$& $0.54\pm	0.08$& $3.60^{2.59}_{1.30}$&3.58\\
MACSJ1931-26&	$0.16\pm0.03$&$0.45\pm	0.04$& $2.81^{0.95}_{0.65}$&2.80\\
RXJ1532.8+3021&	$0.11\pm0.05$&$0.34\pm	0.08$& $3.09^{3.90}_{1.46}$&3.07\\
MACSJ1720+3536&	$0.22\pm0.06$&$0.53\pm	0.08$& $3.40^{1.40}_{0.80}$&3.39\\
MACSJ0429-02&	$0.19\pm0.11$&$0.53\pm	0.13$& $2.78^{5.46}_{1.45}$&2.77\\
MACSJ1206-08&	$0.25\pm0.08$&$0.60\pm	0.11$& $2.40^{1.77}_{0.91}$&2.39\\
MACSJ0329-02&	$0.20\pm0.06$&$0.50\pm	0.09$& $2.50^{1.71}_{0.92}$&2.48\\
RXJ1347-1145&	$0.31\pm0.13$&$0.79\pm	0.19$& $2.54^{2.89}_{1.18}$&2.53\\
MACSJ1311-03&	$0.14\pm0.02$&$0.32\pm	0.19$& $2.28^{1.96}_{1.47}$&2.27\\
MACSJ1423+24&	$0.18\pm0.08$&$0.41\pm	0.06$& $2.27^{2.42}_{0.93}$&2.26\\
MACSJ0744+39&	$0.20\pm0.03$&$0.49\pm	0.04$& $2.44^{0.66}_{0.49}$&2.43\\
CLJ1226+3332&	$0.43\pm0.07$&$1.08\pm	0.09$& $2.51^{0.73}_{0.53}$&2.50\\
\hline
\end{tabular}
}
\end{table}

\section{Conclusion}
In the recent years, the search for detecting the possible deviations from General Relativity, has become an active area of research. Several tests have been proposed to test the validity of GR and/or mark the possible discrepancy between the predictions of GR and the observations. 

Here we analyzed the sparsity parameter in the context of $\Lambda$-gravity and compared the results with the reported data. The analysis shows a difference between the predictions of the parameter in the both contexts. However, the predictions of sparsity parameter in the context of $\Lambda$-gravity are fully consistent with the observational data, as in all cases it lies in the error limits of reported values. As a result, we can expect that together with gravitational lensing \cite{GS3}, the sparsity parameter can be considered as another useful test to mark the direct prediction of $\Lambda$-gravity in the future.

\section{Acknowledgments}

AS acknowledges the partial support by the ICTP through AF-04.

\end{document}